\documentclass[11pt,twoside]{article} \usepackage[margin=1in]{geometry}
 \newcommand\aut{Leonid A. Levin}
 \newcommand\ttl {Internal classes and external sets in math foundations}
 \usepackage {amsmath,amssymb,amsthm,bm,mmap,refcount}\usepackage{microtype}
 \usepackage[T1]{fontenc} \usepackage[unicode]{hyperref}
 \AtBeginDocument{
  } 
\begin{document}

\newcommand\hreff[1] {\href{https://#1}{https://#1}}
\newcommand\edf{{\raisebox{-2pt}{$\,\stackrel{\mbox{\tiny df}}=\,$}}}
\newcommand\p\paragraph \newcommand\trm[1]{{\boldmath\bf\em #1}}
\newcommand\ov\overline \newcommand\un\underline\frenchspacing
 \newcommand\then\Rightarrow\newcommand\ex\exists \newcommand\all\forall
\newcommand\0\emptyset\newcommand\stmn\setminus \newcommand\IP{{\bf IP}}
\newcommand\rn[1]{{\mathbf R^\l_{#1}}}\renewcommand\O{{\mathbb O}}
\newcommand\m{{\mathbf m}}\newcommand\T{{\mathbf T}}
\newcommand\I{{\mathbf I}}\newcommand\W{{\mathbf\Omega}}
\newcommand\PH{{\bf P\raisebox{2pt}{$\boldsymbol\chi$}}}
\newcommand\R{{\mathbb R}}\newcommand\Q{{\mathbb Q}}\renewcommand\r\rho
\newcommand\N{{\mathbb N}}\newcommand\Z{{\mathbb Z}}\newcommand\g{\gamma}
\renewcommand\l\lambda \renewcommand\a\alpha \renewcommand\b\beta
\renewcommand\d{{\mathbf d}} \newcommand\K{{\mathbf K}}
\newcommand\tm[1]{{\mathbf t_{#1}}} \newcommand\IM{{\bf IM}} \catcode`\"=12
\newcommand\sdf{\oplus}\renewcommand\sdf{\triangle}
\newcommand\M{{\mathbf M}}\newcommand\vfi\varphi
\newcommand\D{\Sigma}\newcommand\DD{\D^{\mathbf{\un0}}}
\newcommand\rst[1]{{\boldsymbol{\wedge}_{#1}}}\newcommand\bt{\mathbf2}
\newcommand\ang[1]{{\langle{#1}\rangle}}

 \newcommand\urladdr{www.cs.bu.edu/fac/Lnd} \title
 {Set theory in the foundation of math;\\ internal classes and external sets}
 \author {\aut\thanks {\address.} ~(\hreff{\urladdr})} \newcommand\address
 {Boston University, CAS, 665 Commonwealth Avenue, Boston, MA 02215}
 \date{} \maketitle

\begin{abstract} Usual math sets have special types: countable, compact,
open, occasionally Borel, rarely projective, etc. Each such set is
described by a single set theory formula with parameters unrelated to
formulas. Exotic expressions involving sets related to formulas of
unbounded quantifier depth appear mostly in esoteric or foundational
studies. Recognizing the internal to math (formula-specified) and external
(parameter-based) aspects of math objects greatly simplifies foundations.

I postulate that external sets (not internally specified, constituting
the domain of quantifiable variables) are hereditarily countable and
independent of purely formula-defined classes, i.e. with finite
algorithmic information about them. Variables for classes are not
explicitly quantified.

This opens a way to eliminate all non-integer quantifiers in set theory
sentences. The restrictions seem to require almost no changes in math
papers, only reinterpreting some formalities.
 \end{abstract}\maketitle

\section {Introductory remarks: the problem.}

 I always wondered why math foundations as viewed by logicians are so
distant from the actual math practice. For instance, cardinality theory
-- the heart of ZFC set theory -- is almost never used beyond looking at
which sets are countable and which are not. I see the culprit in the
blurred distinction between two types of collections different in nature.

One type is pure classes, defined by math properties. The really useful
math properties involve very limited quantifier depth. Yet, provisions
for open-ended hierarchies are motivated by perceived need to handle
objects not defined by readily envisioned properties. Indeed, math needs
to handle external objects, not defined by any math properties at all,
for instance, random sequences. These \trm {externals} constitute the
other type: math handles them but generally does not internally
specify.\footnote
 {In reality, external data end up finite. Yet, infinities are neat.
 Even for finite objects their termination points are often ambiguous
 and awkward to handle. As $0$ is a great simplifying approximation
 to negligible $\varepsilon$, so is $\infty$ for $\frac1\varepsilon$.
 And note that the space $\ov\R$ of infinitely long reals is compact,
 having more "finite-like" features in that than $\Q$.}

\subsection {Self-referentials and cardinalities.} Cantor's axioms asserted
 that all set theory formulas define (quantifiable) sets. The effect
was formulas with quantifiers over formulas. This self-referential
 aspect turned out to be fatal. Zermelo and Fraenkel reduced this aspect
by (somewhat {\em ad hoc}) restrictions on how Cantor's axioms treat
cardinalities: \mbox{Replacement} axiom preserves cardinality bounds of
 sets; only a separate Power Set axiom increases them.
 This cardinality focus has questionable relevance. Math papers
 almost never look at distinctions in uncountable cardinalities.

And there is something exotic for mathematics in the Power Set axiom.
Usual math sets have special types: countable, compact, open,
occasionally Borel, rarely projective, etc.\\ Generic subsets from Power
Set classes, with no descriptions, find little use in mathematics.\footnote
 {Admittedly, the sets from the rarely used axiom of choice may pose
  an issue here. But even they can be, with some stretch, regarded
  as internally specified -- by adding a symbol for (thus canonical,
  but not otherwise described) injection of reals into ordinals.}

And as all sound axiom systems have countable models, cardinalities feel
like artifacts, designed to hide self-referential aspects.
 Many papers in Logic (e.g., \cite{hf,rh,rm}) aimed at isolating math
segments where more ingenious proofs can replace the use of the Power Set
axiom and its uncountable sets. But this breaks the unity of math: a
problem. And complicating proofs is unattractive, too.\\ Reals can be as
tricky as any larger sets. The issue is in self-referentials, not in sizes.

\subsection{Some terminology and notation.} I generally refer to collections
as \trm {spaces} or \trm {classes}, only informally calling them sets.
Formal use of the term \trm {sets} will be restricted to the domain of
quantifiable variables. I use metavariables for classes, but do not apply
quantifiers to them, except for the implicit universal quantifier on all
free variables in an expression. I assume known basics, such as
algorithms, and their set theory~expression.

 For brevity, I may refer to hereditarily finite sets as integers,
 and to hereditarily countable ones as reals or sequences.
 Only such "integers" and some "reals" I formally call sets.
 I describe classes $\ov{T_p}$ by including them in transitive
 collections indexed by "real" parameters $p$, and specified by
 membership formula $T_p(q)$ meaning $\ov{T_q}{\in}\ov{T_p}$.
 Ordinals $o\,{\in}\,\O$ are transitive ($\bigcup o\subset o$) sets totally
ordered by membership. Natural numbers $i\in\N$ are finite ordinals.
 It is convenient to include an ordinal $o(p)$ in the indexing parameter
$p$ and restrict membership formula $F_p(q)$ by $o(q){<}o(p)$.
 Then the Foundation axioms are needed only on (subclasses of) sets. Also,
 $\ov{F_s}{\subset}\ov{F'_t}{\subset}\ov{F_s}\then\ov{F_s}{=}\ov{F'_t}$,
by Extensionality axioms. So for $F{=}F'$, it can be required to have a
clause $\all p,s,t\;((\ov{F_s}{\subset}\ov{F_t}{\subset}\ov{F_s})
\;{\then}\;(F_p(s){\then}F_p(t)))$.

Now, $F{\in}\D^1_*\edf\bigcup_k\D^1_k$ implies ranks $k$ for spaces.
($\edf$ denotes definitions.) Quantifiers bind only sets, not properties
$F$, so apply only to spaces below some rank. \trm {Pure} are classes
$\ov{T_p}$ with empty $p$. Examples are $\W=2^\N,\,\O$, or classes
$F\in\D^0_*\edf\bigcup_k\D^0_k\subset\W$ of integers,
 or $F\in\DD_*\edf\bigcup_k\DD_k$ of reals.

\section {Purging the tricksome objects.}\label{prg}

 Making explicit the internal (formula-defined) and external
(parameter-based) aspects of math objects clarifies their nature, allows
a better focus on issues brought by these similar but different sources.
Pure classes are specific but tricksome.
 Treated carelessly they easily bring paradoxes.

Classes will not be quantified or put in collections with no rank limits:
that would only extend the language of allowed formulas. Quantifiable
variables range over "proper" sets -- the external data. No reason to
expect self-referential issues with variables covering only external
objects, unrelated to math properties. Such issues arise from math
properties on variables ranging over objects defined themselves by
properties. Externals may be chaotic, but not really tricksome.

Those sets can be put in collections based on properties and relations
with other sets. This forms general math objects: spaces, specified by
Set Theory formulas with some free variables taken for external
parameters. They carry both chaos and tricks, but only tricks based
on single formulas.

Using information from formulas themselves in their parameters is
redundant and would not increase the expressive power of the formulas.
 Besides, how could external parameters acquire infinite uncomputable
information on formula-defined classes? Yet, seeing no need, or no
realistic mechanism for such relations is not the same as ruling them out.
 The algorithmic information theory (see sec. \ref{ki}) comes to the
rescue. With sets clearly distinguished from pure classes it allows a
radical insight.

\subsection {Independence Postulate \IP}\label{ip}

 Parameters may be infinite not only in length (like, say, the number
$\pi$), but also in complexity.\\ Yet, even then, they can have only finite
dependence on pure classes. This is formalized as finite (small, really)
mutual algorithmic information between them. The "physical" meaning of
that stems from Independence Conservation Inequalities. They assure that
no algorithmic or random processes, nor any combinations of them,
increase the amount of algorithmic information their inputs have about
any specific sequence, such as one defined by a formula. (\cite {fi} has
more discussion.)

This justifies a powerful family {\IP} of axioms:
 $\I(\a{:}\,\ov F)<\infty$. It prohibits the externals with infinite
information about any pure class $F\in\D^0_*$. This family opens a way
to eliminate all non-integer quantifiers from set theory sentences.
 Note, some sequences are both generatable (thus legitimate externals)
 and definable by formulas. But by {\IP} they have finite information
about themselves, thus all are computable (a sort of Church's Thesis).

So, {\IP} conflicts with the Replacement axioms for sets.
 (For spaces, these axioms are merely definitions.)
 Thus I restrict the Replacement axioms to just one: "Classes with
membership graphs computably enumerable {\bf from sets} are sets."
 Note too, {\IP} excludes $\a{\in}F{\in}\DD_*$ unless $\a$ in $F$
reduce to a positive fraction of $\W$. (This condition uses only integer
\mbox{quantifiers}.)

But the reverse does not follow from \IP. So, it needs support from more
axioms.\\ One is the Primal Chaos axiom (\PH):
 {\em "Each set reduces to (its membership graph is enumerable from)
 even-indexed digits of some Kolmogorov--Martin-Lof random $\b{\in}\W$."}
 (For classes, i.e. in ZFC, it is Gacs-Kucera theorem: see, e.g., \cite{gk}.)
 Note, a {\bf random} sequence respects the {\IP} if and only if it is
\trm {generic}, i.e. is outside of all arithmetic classes
$X\in\DD_*$ of measure zero.

\subsection {Models.}\label{mdl}
 Our axioms are ZFC-consistent, having, as all consistent theories do,
countable models.\\ Each model is a reducibility \trm {ideal}.
 This means that with each pair of sequences it includes a sequence they
both are computable from and also all sequences computable from them.

A countable model has a reduction basis: a chain of sequences $\g^{(k)}$,
each computable from $\g^{(k+1)}$, and such that all sequences in the
model are computable from them. Our axioms reduce each of them to a
generic sequence (and the basis can have these reductions reversible).

We can represent such a basis as a set of \trm {trims} of a combined
sequence $\g$. Trims are obtained by dropping a fraction of its digits,
say, those indexed by multiples of $2^k$: dropping every second digit, or
every fourth, eighth, etc. Call such models \trm {internal} if $\g$ is
generic itself.

Internal models respect a family {\IM} of axioms reducing all
quantifiers to integer ones.\\ These axioms assert, for almost all $\g{\in}\W$,
the equivalence of each closed formula $F$ to a modification $F'$ of it
that replaces all 2nd order variables $\a_i$ by algorithms
$A_i(\g^{(k_i)})$ computing those sequences from trims of that $\g$, with
$k_i$ and algorithms $A_i$ quantified as integers. {\em Cf.} \cite{L76}.

\section {Algorithmic information: basic concepts.}\label{ki}

 Here $\|x\|\edf n$ for $x{\in}\{0,1\}^n$ and
 $\|t\|\edf\lceil\log_2t\rceil{-}1$ for $t{\in}\ov{\R^+}$, $\a_{[i]}$ is $\a$'s
 $i$-bit prefix.\\ $\mu(T)$ is the mean of a distribution $\mu$ on
 $T{:}\,\W{\to}\,\ov\R$. $\l(x)\edf2^{-\|x\|}\!$ is the uniform measure.

\trm {Partial continuous transformations} (PCT) on $\W$ may fail to
narrow down the output to a single sequence, leaving a compact set of
eligible results. (Singleton outputs $\{\b\}$ are interpreted as
$\b\in\W$.) So, PCT graphs are compact sets $A\subset\W{\times}\W$ with
$A(\a)\edf\{\b:(\a,\b){\in}A\}\ne\0$.

\trm {Preimages} $A^{-1}(s)\edf\{\a:A(\a)\subset s\}$ of open
$s\subset\W$ are always open.

\trm {Closed} PCTs also have closed preimages of all closed $s$.

\trm {PCT output distributions} $\mu(T)\edf\l(T_A)$,
 $T_A{:}\,\a\mapsto\inf T(A(\a))$ on $\l$\\ are \trm {semimeasures},
 i.e. superlinear: $\mu(a^2T+S)\ge a^2\mu(T){+}\mu(S)$, $\mu(\W){\le}1$.

\trm {Computable} PCTs have algorithms enumerating the clopen subsets of
$\W^2{\stmn}A$.\\ I generally assume PCTs to be such.
 A universal PCT $U(p\a)$ on prefixless $p$\\ computes $A_p(\a)_{[n]}$
in $\tm{p\a}(n)$ steps (and $U(p){\in}\N$ in $\tm p$ steps).

A \trm {computably enumerable (c.e.)} by algorithm $\ang t$ real function
$t$ is\\ the $\sup$ of a c.e. family of basic continuous ones in an
ordered Banach space.

\trm {Dominant} in such a space is a c.e. $f$ in its unit ball $B$
 if all c.e. $g$ there are $O(f)$.\\ Such is $\sum_ig_i/(i^2{+}i)$
 if $(g_i)_i$ is a c.e. family of all c.e. $f$ in $B$.

\trm{Complexity $\K(x)$} is $-\|\m(x)\|$ for $\m$ dominant in $l^1$,
 $\sum_x\m(x)<1$.\\ Let $\bt_x(\a)=2$ if $\a{\in}x\W$, else ${=}0$.
 $\M(x)\edf\M(\bt_x)/2$ is a dominant c.e. semimeasure.

\trm {$\mu$-rarity $\d_\mu(\a)$} is $\|\lceil\T_\mu(\a)\rceil\|$;
 $\d_\mu^\vfi\edf\|\lceil\T_\mu^\vfi\rceil\|$,
 $\T_\mu\edf\sum_i\{T_i{:}\,\mu(\T_i){\le}\m(i)\}$,\\
 c.e. family $(T_i)_i$ of all c.e. $T$ on $\W$. Its $\vfi$-mean
 $\T_\mu^\vfi$ is $\sum_i\{\vfi(T_i){:}\,\mu(\T_i){\le}\m(i)\}$.\footnote
 {The earliest complexity expression of Martin-Lof test $T$ was equal to
 $\sup_{t,x}\{\bt_x^t{:}\,\mu(\bt_x^t){\le}\M(x)\}$.\\ But its $\mu(T)$
 diverges unless $\M(x)$ changed to $\M(t,x)$. In general, call a c.e.
 family $t$ of c.e. $t_x{:}\,\W{\to}\R^+$ a tree if $t_x t_y{=}0$\\
 or $t_x{\ge}t_y$ or $t_y{\ge}t_x$ for all $x,y$. Let support of
 $t_x$ be $\widehat{t_x}\,\W$. Then extend $T$ to all trees $t$ as
 $\sum_t\sup_x\{t_x{:}\;\mu(t_x){\le}\M(\ang t,\widehat{t_x}\,)\}$.\\
 Even more general is to randomize $t$ giving it coin flips $\r{\in}\,\W$
 and averaging $T$ over $\r$.}

 \trm {Kolmogorov-Martin-Lof (KML) $\l$-random}
 are $\a\,{\in}\,\rn{}\edf\rn\infty$, for
 $\rn c\edf\{\a{:}\,\d_\l(\a)\,{<}\,c\}$.\\ $\rn{}$ consists of
 all $\g$ with $\sup_i{\M(\g_{[i]})}/{\l(\g_{[i]})}<\infty$.

 \trm {Mutual information (dependence)}
 $\I(\a_1\,{:}\,\a_2)$ is $\d_{\M\otimes\M}((\a_1,\a_2))$.

\subsection {Some useful properties of randomness: a brief sketch.}
 \begin{description} \item [Conservation:]
 $\d^{A(\vfi)}_{A(\mu)}{\le}\,\d^\vfi_\mu{+} O(1)$. Indeed, let $i'{=}j$
 where $T_j{=}A(T_i)$. $\sum_i\{A(\vfi)(T_i){:}\,A(\mu)(\T_i){\le}\m(i)\}{=}$
 $\sum_i\{\vfi(A(T_i)){:}\;\mu(A(\T_i)){\le}\m(i)\}=$
 $\sum_i\{\vfi(T_{i'}){:}\;\mu(\T_{i'}){\le}\m(i)\}=$
 $O(\sum_j\{\vfi(T_j){:}\;\mu(\T_j){\le}\m(j)\})$.
 \item [Soundness:] $\d^\mu_\mu=0$ is straightforward.
 \item [Triviality for $\M$:] $\d_M=O(1)$. Indeed: Let $f{\circ}\mu$ have
 $\mu(gf)$ mean on $g$.\\ Take a c.e. family $\M_i$ of c.e. semimeasures
 such that $\M_i=T_i{\circ}\M$ if $\M(T_i)\le\m(i)$.\\ Then c.e.
 $\M'\edf\sum_i \M_i= \Theta(\M)$. Let $\rst x(f)\edf\inf_{\a\in x\W}f(\a)$.
 Then $\M(x)\rst x \le {\bt_x}{\circ}\M/2$,\\
 $\M(x)\T_\M^{\rst x} = \T_\M^{\M(x)\rst x} \le\T_\M^{{\bt_x}\circ\M/2}\le$
 $\M'(x) = O(\M(x))$. So, $\T_\M^{\rst x}=O(1)$.
 \item [Completeness of \IP:]
 $\M(F){=}0\iff\ex\a\all\b{\in}\,F\;\I(\a:\b){=}\infty$. Indeed:
 Let $F\subset\limsup F_i$ for\\ clopen $F_i$ such that $\M(F_i)<2^{-i}$.
 Then $\a(i)\edf(F_i,\K(F_i))$ gives $\I(\a:\b)=\infty$ for $\b\in F$.
 \end {description}

{\bf Measuring} spaces. Each $F{\in}\DD_*$ has clopen $F_i$,
$\M(F_i{\sdf}F){<}2^{-i}$, $F{\subset}\limsup F_i$.\\ If $F\subset\rn{}$,
 $\l(F)=0$ then $\M(F_i)=O(2^{-i})$, thus $\I(\g{:}\,(F_i)_i)=\infty$
for all $\g\in F$.\\ So, any $\DD_*$ space invariant
 under single-digit flips includes all sets $\g{\in}\rn{}$ or none.

\section {Weak truth-table reductions to generic sequences.}
\label{wtt}

 {\PH} generates all $\b{\in}\W$ from $\a{\in}\rn{}$. PCT $U(\a)$ can
 use unlimited segments $\a_m$, dropping nearly all their information,
 to produce small segments $\b_n$. But {\IP} makes {\PH} equivalent
 to its stronger form, requiring a {\bf closed}\footnote
 {Yet, by \cite{vv}, $U$ cannot be made total, nor
 reversible on $\a$: some loss unavoidable.}
 PCT $u$, with $m{\sim}n$.
 Let $s_t^n\edf\l(\{\a{:}\,\tm\a(n){<}t\})$, $s_\infty\edf\inf_ns^n_\infty$,
 $s_\a\edf\liminf_ns^n_{\tm\a(n)}$. Let $\a$ be generic.
 Then $s_\a{<}s_\infty$, as $\l(\{\a\,{:}\;s_\a{=}s_\infty\}){=}0$.
 Take $r\in(s_\a,s_\infty)$. \newcommand\n{\mathbf n}
 $\tau_r^n\edf\min\{t\,{:}\,s_t^n{>}r\}$ is computable, so take
 a monotone infinite sequence $\n^\a_i$ of all $n$ with $\tm\a(n)<\tau_r^n$.

$\I(\a\,{:}\,0'){<}\infty$, so $\all^\infty i\;\n^\a_i{<}\,\max_{p<i\|i\|^2}t_p$.
 Then $U_c:\W\,{\to}\,\{\#,0,1\}^\N$ avoids divergence by\\ diluting $U(\g)$
 with $\min_{p<i\|i\|^2+c}\,\{\tm p{:}\;\tm p{>}\n^\g_{i+1}\}\le\infty$
blanks $\#$ after $\n^\g_i$-th bits. $U_c(\g)$ carries no extra information
of $\g$ absent in $U(\g),\n^\g$.
 As $U_c$ never diverges, $\mu\edf U_c(\l)$ and its distribution function
$d(r)\edf\mu([\#^\N,r])$ (for $r$ with $\mu(\{r\}){=}0$) are computable.
So $U'(\a)\edf d(U_c(\a))$ maps $\rn{}$ to $\rn{}$ as its outputs'
distribution is $\le\l$. From $\b=U'(\a){\in}\rn{}$ we recover $U_c(\a)$
and $U(\a)$.

The PCT $u(\b)\edf U(\a)$ is closed (a w.t.t. reduction): the
input segments it uses are not much longer (by codes $p$ for $\n^\a$
bounds) than the output's.\footnote
 {Viewing $\#^+\{0,1\}^+$ segments of $U_c(\a)$ as integers
 makes $\mu$ (on $\N^{<\N}$ prefixes)\\ a computable distribution
 on (finite and infinite) sequences. $U'$ gives them short codes.}

\section {Elimination of 2nd order quantifiers with \IP}\label{ia}

In a sentence, with no free variables, all non-integer quantifiers can be
 eliminated.\\ This uses the Internal Models axioms ({\IM} of Sec.\ref{mdl}).
 Yet, adding the entire {\IM} family strikes me as less elegant or
intuitive than a single neat axiom, such as \PH. It seems tempting to try
such elimination with free variables, too, by induction on the number of
quantifiers, employing much simpler axioms, or just {\PH} itself.
 However, this path is barred by an obstacle:

 Gacs-Kucera theorem (\PH) is consistent with {\IP}.\footnote
 {In fact, under {\IP} it is a special case of an older
 Prop. 3 of \cite{L76}, Prop. 4 of \cite{L84}.}
 But its relativization to $\a$-algorithms with a free variable $\a$ used
 as an oracle, is not:
 Call a PCT $A$ (recursively) \trm {one-way}
 on $X{\subset}\rn{}$ if $\l(X){>}0$, $A(X){\subset}\rn{}$,
 and no randomized $B$ invert it, i.e. $\l(Y){=}0$
 for $Y{=}\;\{(\a,\r){:}\;\b\edf B(\a,\r){\in}X,A(\b){=}\a\}$.

 {\bf Example 1} (follows from \cite{bz}). Take $A(\b)\edf\a$ where
 $\a(i){=}\b(h(i))$.

 If algorithm $h$ has undecidable $h(\N)$,
 it cannot be resolved by seeing bias in $\b(j)$ distribution.
 Gacs-Kucera theorem fails if relativized to algorithms running with
oracle $\a{=}A(\b)$. Then $\b\in X$ do not $\a$-reduce to reals $\r$ random
under $\a$-c.e. rarity. Otherwise call $\a{\in}\rn{}$ (not in any $A(X)$)
\trm {primal}. Also, call $\b_1,\b_2$ $A$\trm{-siblings} if they differ
in infinitely many digits and $A(\b_1){=}A(\b_2)$.\\
 Taking hypersimple $h(\N)$ in Example 1, makes all $A$-sibling-free
classes to have measure 0, and\\ the class of all pairs of siblings, too:
 Siblings differ on infinite subsets of hyperimmune $\N\setminus h(\N)$,
 thus have 0 chance to be generated. So, no pairs of siblings are sets.
 % T.14 doi.org/10.1016/j.apal.2010.09.007 arxiv.org/abs/1009.5894

 Thus, $\ex\b\,P(\b)\&A(\b){=}\a$ depends on which of the siblings
 mapped to $\a$ exists.\\ So, $\ex\b$ quantifier is model-dependent and
 cannot be eliminated.

 The elimination of non-integer quantifiers
 can work only for closed sentences, not via the mentioned induction.
 This leaves open the intriguing question of replacing {\IM} axioms
 with something more intuitive. The {\PH} axiom may need strengthening,
 e.g., requiring all reals to reduce to primal, not just random, reals.
 It may need to assure that primal $\a$ can be combined with any
 other $\b$ into another primal $(\a,\g)$ to which $\b$ reduces.
 Primality, unlike randomness, involves real quantifiers.

\subsection {Wishful thinking.}

 The above questions seem an interesting area to think about.
 And {\IM} may turn out even be theorems.
 Any sentence $G$ consistent with \IP, {\PH} holds in a countable
 model M$_G$ of the axioms. Let such $G$ be $F\,\&\,\widetilde F$
 for $\widetilde F\edf\l(\{\g{:}\,F'(\g)\}){=}0$. Then M$_G$
 respects {\PH}, {\IP} with some $\Psi{\in}\,\D^0_*\cap\rn{}$,
 fixed $c$, $\rn\Psi\edf\{\g:(\Psi,\g)\in\rn c\}$ such that
 $\all\g\,((\widetilde F\,\&\,F'(\g))\then\g{\notin}\,\rn\Psi)$.
 For a contradiction we need to express M$_G$ in an "internal style",
 enumerating its sets from trims $\g^{(k)}$ of $\g\in\rn\Psi$.
 If a reduction basis of M$_G$ is Turing-equivalent to $\a_i{\in}\rn\Psi$,
 we need to merge all $\a_i$ into such $\g$ as trims.

\section {Some discussion.}

\vspace{-7pt}Cantor's axioms' license for formula-defined sets led to fatal
consistency problems. Zermelo-Fraenkel (somewhat {\em ad hoc})
cardinality-based restrictions diffused those but left intact their
self-referential roots. This seems just an example of uncoiling a vicious
circle into a vicious spiral.

 The result was a Babel Tower of cardinalities, other hierarchies
finding little relevance in math. And the generality of these
hierarchies is illusory. Expanding set theory with more formula types,
axioms, etc. has no natural limit. Benefits are few and eventual
consistency loss inevitable: \cite{fi}.

{\bf A clean way out may be recognizing the distinction between\par
collections internal-to-math, specified by its formulas, and externals\par
that math handles as values of variables, without specifying.}

\noindent Internal collections have a limited hierarchy: the allowed
formula types are clear-cut.\\ Any extension would make a new theory,
with its own \mbox{limits}. External objects are independent of internal
ones: have finite information about them. Complexity theory allows to
formalize that, justify the validity for "external data", and use that
for simplifying math foundations.

General math spaces are collections specified by formulas with external
sets as parameters. Collections of them are represented by collections of
those parameters. So, any such space would rely on a single formula, not
on all of them, thus excluding back-door extensions of formula language.

The uniform concept for all collections of math objects with no explicit
types hierarchy is alluring but illusory. The hierarchy of ever more
powerful axioms, models, cardinals remains, if swept under the rug.
Having it explicit and matching math relevance sets a path to simpler
foundations.

 What is left out? -- "Logical" sets, related to infinite hierarchies
 of formulas, such as\\ "the set of all true sentences of Arithmetic".
 Many interesting results on cardinalities and\\
 other hierarchies are internal to Logic.
 More clarity on which issues in Logic are the concerns\\ of foundations
 of mathematics and which are based on its own pursuits may add insight.\\
 All research areas serve each other, yet have their internal interests, too.

 {\bf Acknowledgment.} I am grateful to Albert Meyer for helpful criticism.

\vfill\appendix \section {Some Other {\IP} Applications. ZFC Axioms.}

\subsection {Foundations of probability.} (See \cite{L84} Th.2.)
 Paradoxes in its application led to the K-ML randomness concept $\rn{}$.\\
 {\IP} clarifies its use: For any $S{\subset}\,\W$: $\l(S){=}0$ iff
 $\ex\sigma\; S{\cap}\rn{}\subset\{\g{:}\;\I(\g{:}\,\sigma){=}\infty\}$.

\subsection {Goedel Theorem loophole.} \href
{https://www.marxists.org/reference/subject/philosophy/works/at/godel.htm}
{Goedel writes:}\begin{quote}
 "It is not at all excluded by the negative results mentioned earlier
that nevertheless\\ every clearly posed mathematical yes-or-no question is
solvable in this way.\\ For it is just this becoming evident of more and
more new axioms on the basis\\ of the meaning of the primitive notions that
a machine cannot imitate."\end{quote}
 {\bf No way !} Let a predicate $P$ on $\{0,1\}^n$ extend "proven/refuted"
 partial predicate of Peano\\ Arithmetic. Let $r_n$ be the n-bit prefix
 of a c.e. real $r=\min\rn0$. Then $\I(P:r_n)= n\pm O(\log n)$.\\
 (See \cite{fi}.) No way to obtain such $P$ for any significant $n$,
 by formal or informal methods!

\subsection {ZFC axioms.}

ZFC axioms are sometimes given to undergraduates in an unintuitive, hard
to remember list.\\ Setting them in three pairs seems to help intuition.

Sets with a given set theory property $F$ (possibly with parameters $c$)
are said to form a \trm {class} $\{x:F_c(x)\}$. They may or may not form
a set, but only sets are the domain of ZFC variables.

\begin{enumerate}\parskip0pt
 \item {\bf Membership chains: sources, sinks} (1b anti-dual to 1a):
 \begin{enumerate}
 \item {\bf Infinity.} (No "$\in$"-source in $S$):\hfill\fbox
  {$\ex S{\ne}\emptyset\,\all x{\in}S\ex y{\in}S(x{\in}y)$}
 \item {\bf Foundation.} (Each set has sinks: members disjoint with it):
 \hfill\fbox {$\neg\,\ex S{\ne}\emptyset\,\all x{\in}S\ex
y{\in}S(y{\in}x)$}
 \end{enumerate}

 \item{\bf Sets with formula-defined membership:}
 \begin{enumerate}
 \item {\bf Extensionality.} (Content defines sets):
 \hfill\fbox {$x{\supset}y{\supset}x{\in}t\then y{\in}t$}
 \item {\bf Replacement.}\footnote {An axiom $R_F$ for each class
 $F_c(X)\;\edf\;\{y:\ex x{\in}X\,F_c(x,y)\}$.}
 (Classes with cardinality bounds are sets):\\\null\hfill\fbox
 {$(\all x\,\ex Y{\supset}\,F_c(\{x\}))\,\then$
  $\all X\,\ex Y{\supset}F_c(X){\supset}Y$}
 \end{enumerate}

 \item{\bf Functions inverses:} $f^{-1}\edf\,\{g\subset f^T:f(g(f(x)))=f(x)\}$:
 \begin{enumerate}
 \item {\bf Power set.} ($f^{-1}{\subset}G$ is a set, take $h{=}f^T$):
  \hfill\fbox {$\all h\ex G\,\all g{\subset}h\,(g{\in}G)$}
 \item {\bf Choice.}\footnote {Feasibility of computing
  inverses is a dramatic open problem in Computer Theory.}
  ($f^{-1}$ is not empty): \hfill\fbox {$\all f\,\ex\,g\in f^{-1}$}
 \end{enumerate}\end{enumerate}

\vfill\subsection {The ZFC modifications discussed above include:}

\begin{enumerate}\itemsep2pt\parskip0pt
 \item Restrict Replacement to computable $F$. Add its opposite, the \IP.
 \item The Power Set replaced with its opposite: "All sets are countable."
 \item Add Primal Chaos axiom. (A question if something stronger is
needed is open.\\ At worst, it could be the Internal Model axioms family.)
 \item Restate Foundation and Extensionality as families to cover classes.
 \item The Choice may be dropped or replaced with adding to the language\\
a (named, not described) class postulated to inject reals into ordinals.
 \end{enumerate}

Mathematical spaces are classes of sets $q$ satisfying formulas $F$ with
external parameters $p$. Collections of spaces are treated as collections
of those parameters. Quantifiers bind parameters, not properties $F$.
 Papers may have families of theorems parameterized by formulas.\\ Such are
used now for the so-called Large Categories that are not Zermelo-Fraenkel
sets.\\ They are also used for families of axioms, like Induction axioms.

We handle such families by allowing variables for classes but not quantifiers
 on them, except\\ for an implicit universal quantifier over all free
variables in an expression. So parameterized will\\ be the Foundation and
Extensionality axioms, because they must apply to spaces, not just sets.

\end{document}